\documentclass[aps,prb,twocolumn,groupedaddress,10pt,nofootinbib,a4paper,notitlepage, showpacs]{revtex4-1}

\usepackage{dcolumn}% Align table columns on decimal point
\usepackage{bm}% bold math
\usepackage{amsmath}
\usepackage{amssymb}
\usepackage{amsfonts}
\usepackage{amsthm}
\usepackage{amscd}
\usepackage{color}
\usepackage{graphicx}

\def\be{\begin{equation}}
\def\ee{\end{equation}}

\def\bra#1{\mathinner{\langle{#1}|}}
\def\ket#1{\mathinner{|{#1}\rangle}}

{\catcode`\|=\active
  \gdef\set#1{\mathinner{\lbrace\,{\mathcode`\|"8000\let|\midvert #1}\,\rbrace}}
  \gdef\Set#1{\left\{\:{\mathcode`\|"8000\let|\SetVert #1}\:\right\}}}
\def\midvert{\egroup\mid\bgroup}
\def\SetVert{\egroup\;\mid@vertical\;\bgroup}

\newcommand{\mbf}[1]{\mathbf{#1}}

\begin{document}

% \title{The absence of noise-assisted transport in the presence of certain initial correlations }
\title{Avoiding dark states in open quantum systems by tailored initial correlations} 
\author{P. Schijven}
\email{Email: petrus.schijven@physik.uni-freiburg.de}
\author{O. M\"ulken}

\affiliation{Physikalisches Institut, Universit\"at Freiburg, Hermann-Herder-Strasse 3, 79104 Freiburg, Germany}
\date{\today}


\begin{abstract}
We study the transport of excitations on a V-shaped network of three
coupled two-level systems that are subjected to an environment that
induces incoherent hopping between the nodes. Two of the nodes are coupled
to a source while the third node is coupled to a drain. A common feature
of these networks is the existence of a dark-state that blocks the
transport to the drain. Here we propose a means to avoid this state by a
suitable choice of initial correlations, induced by a source that is
common to both coupled nodes. 
\end{abstract}

\maketitle

\section{Introduction}
The transport dynamics of energy or charge 
%excitations and/or electrons 
in many physical
systems can be described by considering transport on a network of coupled
2-level systems. Depending on the physical system, the transport can
range from being purely coherent to being purely incoherent.
In the former case the dynamics can
be described by continuous-time random walks
\cite{VanKampen1990}, 
while in the latter case the dynamics follows from Schr\"odinger's
equation, which for complex systems and certain choices of the Hamiltonian
defines so-called
continuous-time quantum walks
\cite{Farhi1998,Muelken2011}. 
Coupling of a quantum system with purely coherent dynamics to a bath of
harmonic oscillators (for instance phonons) can lead to a mixture of
coherent transport and incoherent hopping induced by the environment.
The dynamics (of the reduced density matrix) can be described by a quantum
master equation of Lindblad type \cite{BreuerOpenQS}, where a certain choice of Lindblad
operators defines so-called
quantum stochastic walks \cite{Whitfield2010}. 

In this work we consider transport on a V-shaped trimer configuration,
resembling for instance a system of coupled quantum dots
\cite{Dominguez2011, Michaelis2006, Groth2006}. The transport will be generated by
connecting a (incoherent) single source to the end nodes of the V-shaped trimer and a 
(incoherent) drain to the middle node, see below.
Most theoretical works have focused on the
case where the source creates an excitation on a single node of the
network \cite{Caruso2009,Sarovar2011}. One can also, however, consider
the possibility where a source is connected to multiple nodes of the
network. Such a source can then create an excitation that is in a
superposition of these nodes, leading to initial correlations between
these nodes.

A common feature of these V-shaped, or circular, trimer configurations is
the existence of a dark state \cite{Michaelis2006,Emary2007,Brandes2005,
Molmer1993, Caruso2009, Agliari2010}. Such a state causes the excitation to become
trapped in the network and therefore leads to a blocking of the transport
in the purely coherent case.
To overcome this problem, one can either introduce an energetic disorder
on the nodes \cite{Emary2007,Caruso2009} or couple the system to a
suitable environment where the decoherence process destroys the
interference effects that lead to the dark state
\cite{Caruso2009,Dominguez2011}. Here we propose a third method: a
suitable choice of initial correlations that are induced by a single
source which is coupled to the end nodes of the network creates an initial
state which is orthogonal to the dark state. This then causes the absence
of the dark state in the transport even in the purely coherent case,
leading to a complete transport to the drain. 
% We then study the influence of the coupling to the
% environment on the transport efficiency and also compute the transport
% properties for other forms of initial correlations and compare our results
% to the case where there are none.

The paper is organized as follows. In Sec. \ref{sec:two} we introduce our
model and provide a detailed discussion on the exact mathematical implementation of a
source that induces initial correlations. In Sec. \ref{sec:three} we
discuss the transport efficiency of different initial configurations with the
help of both analytical and numerical computations.

\section{Modelling the transport}\label{sec:two}
\subsubsection{Coherent and incoherent dynamics}
We consider the dynamics of excitations on a trimer network. The coherent
(quantum) dynamics on this network is described by the 
Schr\"odinger equation, or equivalently by the Liouville - von Neumann
equation, with the general Hamiltonian
%continuous-time quantum walk (CTQW). This model is motivated by the similarities between the master equation of a continuous-time random walk (CTRW) and the Schr\"odinger equation for finite level systems. It is defined by identifying the Hamiltonian $\mathbf{H}_0$ of the system with the transfer matrix $\mathbf{T}$ of the CTRW. In general we can write the Hamiltonian $\mathbf{H}_0$ of the CTQW as:
\be
  \mathbf{H}_0 = \left( \begin{array}{ccc}
              E_1 & V_{12} & V_{13} \\
              V_{12}   & E_2 & V_{23} \\
             V_{13}  & V_{23} & E_3
             \end{array}\right).
\ee
Here, $E_k$ is the site energy of node $\ket{k}$ and $V_{kl}$ are the
transfer rates between nodes $\ket{k}$ and $\ket{l}$. Note that for certain
choices of the site energies and the couplings there exists an eigenstate 
$\ket{D} = (\ket{1} - \ket{2})/\sqrt{2}$ of $\mathbf{H}_0$, having only
an overlap with nodes 1 and 2 \cite{Michaelis2006,Emary2007,Brandes2005, Caruso2009}.

Now, if the system is in contact with an external environment, the total
Hamiltonian takes the form $\mbf{H}_{\text{tot}} = \mbf{H}_0 + \mbf{H}_E +
\mbf{H}_{\text{int}}$, where $\mbf{H}_0$ is the Hamiltonian of the
network, $\mbf{H}_E$ is the Hamiltonian of the environment and
$\mbf{H}_{\text{int}}$ specifies the interactions between the network and
the environment. When the environmental correlation time is small
compared to the relaxation time of the system, one can describe the
dynamics on the network by a master equation in Lindblad form
\cite{BreuerOpenQS}:  
\be\label{eq:lindblad}
   \frac{d \bm{\rho}_N(t)}{dt} = -i \left[ \mbf{H}_0, \bm{\rho}_N(t) \right] + \sum_{k,l=1}^{3} \lambda_{kl} \mathcal{D}(\mbf{L}_{kl}, \bm{\rho}_N(t)) ,
\ee
with the constants $\lambda_{kl}\geq 0$ for all $k$ and $l$ and
\be
\mathcal{D}(\mbf{L}_{kl}, \bm{\rho}_N(t)) = \mbf{L}_{kl}^{\phantom{\dagger}} \bm{\rho}_N(t) \mbf{L}_{kl}^\dagger - 
\frac{1}{2}\left\{\mbf{L}_{kl}^\dagger \mbf{L}_{kl}^{\phantom{\dagger}}, \bm{\rho}_N(t) \right\} .
\ee
For our model we assume the Lindblad operators $\mbf{L}_{kl}$ to be given
by $\mbf{L}_{kl} = \ket{k}\bra{l}$. The term in Eq. \eqref{eq:lindblad}
corresponding to the operator $\mbf{L}_{kl}$ models the incoherent
excitation dynamics, induced by the environment, between the nodes $l$ and
$k$ with rate $\lambda_{kl}$. 
% An approach formalizing this idea goes under
% the name of the quantum stochastic walk (QSW) \cite{Whitfield2010}.

In the purely incoherent limit we assume simple hopping dynamics to be
described by a Pauli master equation (for the diagonal elements of
$\bm{\rho}(t)$ only).
The corresponding transition rates $\lambda_{kl}$, for $k\neq l$, can be
phenomenologically estimated with Fermi's golden rule, e.g. $\lambda_{kl}
= \Lambda |V_{kl}|^2$. Here $\Lambda$ is a constant that captures the
particular details of the environment (for simplicity, we assume that
$\Lambda = 1$).  
We further assume additional pure dephasing, which is taken to be
identical for all the nodes of our network, i.e. we have Lindblad
operators $\mathbf{L}_{kk}$ with $\lambda_{kk} = \lambda$ for all $k$
, where $\lambda$ is the global dephasing rate \cite{Rebentrost2009}.
% We further assume that the coupling to the
% environment is identical for all the nodes of our network. Since the
% Lindblad operators $\mathbf{L}_{kk}$ correspond to pure dephasing, see for
% example \cite{Rebentrost2009}, we assume the constants $\lambda_{kk}$ to
% be $\lambda_{kk} = \lambda$ for all $k$, with $\lambda$ the global
% dephasing rate for our system. 

Since the dissipative terms in the Lindblad master equation induce incoherent
hopping between the nodes, we can introduce a parameter $\alpha$, with
$0\leq \alpha \leq 1$, that allows us to interpolate between purely
coherent dynamics ($\alpha=0$)  and purely incoherent dynamics
($\alpha=1$):
\be
\frac{d \bm{\rho}_N(t)}{dt} = (1-\alpha) \mathcal{L}_{\text{coh}}(\bm{\rho}_N(t)) + \alpha \mathcal{L}_{\text{env}}(\bm{\rho}_N(t)) ,
\ee
with $\mathcal{L}_{\text{coh}}(\bm{\rho}_N(t)) = -i \left[ \mbf{H}_0,
\bm{\rho}_N(t) \right]$ and $ \mathcal{L}_{\text{env}}(\bm{\rho}_N(t))
=\mathcal{L}_{\text{incoh}}(\bm{\rho}_N(t)) +
\mathcal{L}_{\text{deph}}(\bm{\rho}_N(t))$. 
This approach is also known as the quantum stochastic walk \cite{Whitfield2010}.
The generator
$\mathcal{L}_{\text{incoh}}$ corresponds to the terms in
\eqref{eq:lindblad} that generate incoherent transfer between the nodes
and the generator $\mathcal{L}_{\text{deph}}$ corresponds to the terms
that generate pure dephasing.
\begin{figure}
 \begin{center}
    \includegraphics[width=0.9\columnwidth]{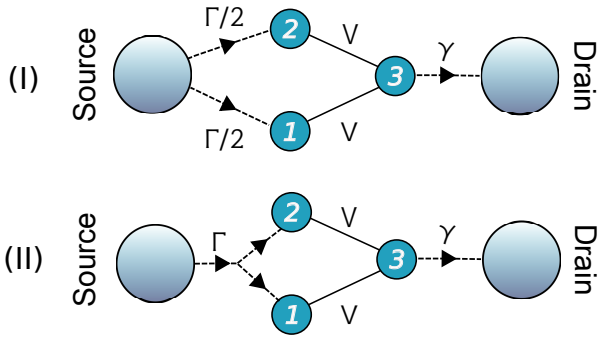}
 \end{center}
 \caption{An illustration of the two different ways of connecting the source(s) to the trimer network: (I) corresponds to 
 the configuration with 2 independent transitions to the network and (II) corresponds to the configuration with a source that creates
 initial correlations between nodes 1 and 2.}\label{fig:trimer-network}
\end{figure}

\begin{figure*}
 \begin{center}
  \includegraphics[width=\textwidth]{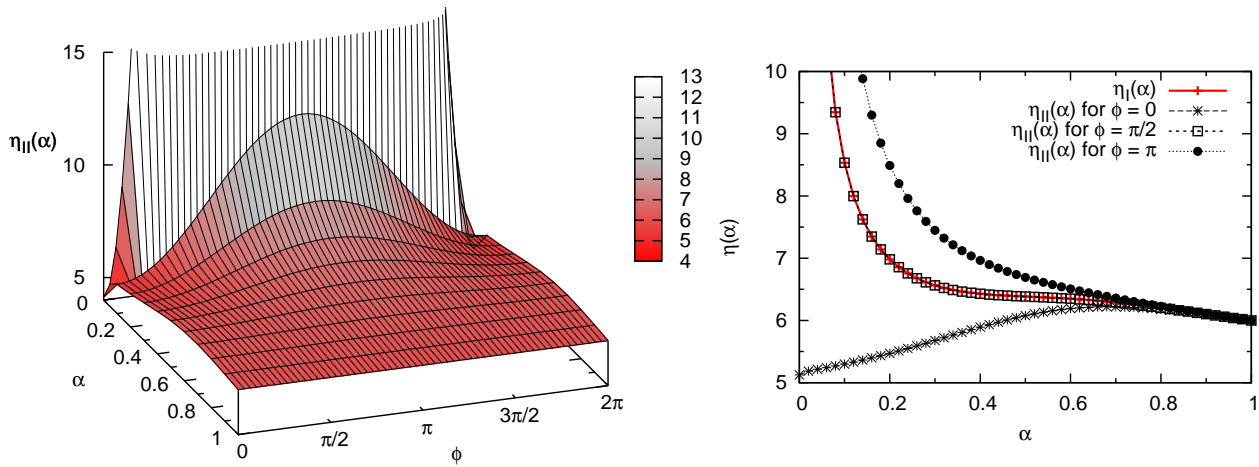}
 \end{center}
 \caption{Dependence of the EST $\eta_{II}(\alpha)$ on the angle $\phi$ for $\lambda = 1$ and $\Gamma = 0.5$. On the left we show
 the general result for all values of $\phi$ and $\alpha$. The right figure shows the cross-sectional curves for $\phi = 0, \pi/2$ and $\phi=\pi$,
 together with the result for $\eta_{I}(\alpha)$.}\label{fig:phi-dep}
\end{figure*}

\subsubsection{Sources and drains}
A source is included in our system as an extra
node $\ket{0}$ that is incoherently
coupled to the nodes of our network in order to make sure that there is only
transport from the source to the network and not back. 
%Then, we can write the new system Hamiltonian as $\mathbf{H} = \mathbf{H}_0 + \epsilon_0 \ket{0}\bra{0}$. 
In
general, transitions from the source to a general state
$\ket{\psi} = \sum_k a_k \ket{k}$ of the network
%. Such incoherent transitions 
can be phenomenologically modelled by the Lindblad operator
$\mathbf{L}_s = \ket{\psi}\bra{0}$, leading to the following additional
term to the master equation:
\be\label{eq:gen-source}
  \mathcal{L}_{\text{source}}(\bm{\rho}_{SN}(t)) = \Gamma \mathcal{D}( \mathbf{L}_{s}, \bm{\rho}_{SN}(t)),
\ee
with $\Gamma$ representing the rate at which the excitation flows into the
network. 
Note that for a $N$-dimensional network with a source, the reduced
density matrix $\bm{\rho}_{SN}$ is represented by a $(N+1)\times (N+1)$ 
matrix.
For an initial preparation in the source node, i.e.  $\bm{\rho}_{SN}(0)
= \ket{0}\bra{0}$, one can show that the density matrix can be written in
the form \cite{Schijven2011}:
\be
  \bm{\rho}_{SN}(t) = \left( \begin{array}{cc}
                      \rho_{00}(t) & 0 \\
		      0 & \bm{\rho}_{N}(t)
                     \end{array}\right),
\ee
where $\bm{\rho}_{N}(t)$ is the density matrix corresponding to the
network nodes. 

In a similar fashion, we include a drain by coupling the 
state $\ket{N+1}$ to the network. The incoherent transition from a state
$\ket{\psi}$ of the network to the drain with rate $\gamma$ can then be modelled by the
term
\be
  \mathcal{L}_{drain}(\bm{\rho}_{SND}(t)) = \gamma \mathcal{D}(\bm{L}_{d}, \bm{\rho}_{SND}(t)),
\ee
in the master equation, with $\bm{L}_d = \ket{N+1}\bra{\psi}$.
Thus the final reduced density operator $ \bm{\rho}_{SND}(t) = \bm{\rho}(t)$
is represented by a $(N+2)\times(N+2)$ matrix.

\subsubsection{Creating initial correlations with a source}
There are now two interesting ways in which we can connect the source
to the end nodes $\ket{1}$ and $\ket{2}$ of the trimer network:
\begin{enumerate}
\item[(I)] The source can either feed node $1$ with rate
$\Gamma/2$ or node $2$ with rate $\Gamma/2$. We assume these processes
to be \emph{independent} of each other. We can model this with \emph{two}
dissipators representing the two independent processes:
 \begin{eqnarray}\label{eq:uncorrelated}
   \mathcal{L}_{\text{source}}^{(1)}(\bm{\rho}(t)) &=& \frac{\Gamma}{2} \mathcal{D}( \ket{1}\bra{0}, \bm{\rho}(t)) + \frac{\Gamma}{2} \mathcal{D}( \ket{2}\bra{0},\bm{\rho}(t)) \nonumber \\
 &=& \frac{\Gamma}{2}\rho_{00}(t) \left(\ket{1}\bra{1} + \ket{2}\bra{2} - 2\ket{0}\bra{0}\right) .
 \end{eqnarray}
\item[(II)] The source can also feed a superposition state $\ket{\psi}$
between node $1$ and node $2$, which can in general be written as
$\ket{\psi} = \left( \ket{1} + e^{i \phi} \ket{2}\right)/\sqrt{2}$. We can
model this process with \emph{one} dissipator:
 \begin{eqnarray}\label{eq:correlated}
   \mathcal{L}_{\text{source}}^{(2)}(\bm{\rho}(t)) &=& \Gamma \mathcal{D}( \ket{\psi}\bra{0}, \bm{\rho}(t)) \\
  &=& \frac{\Gamma}{2}\rho_{00}(t) \bigg[\ket{1}\bra{1} + \ket{2}\bra{2} - 2\ket{0}\bra{0} \nonumber  \\
  && +  e^{-i\phi} \ket{1}\bra{2} + e^{i\phi} \ket{2}\bra{1}\bigg] .
 \end{eqnarray}

\end{enumerate}
See Fig. \ref{fig:trimer-network} for an illustration of these two configurations. 
The key difference between these two choices is that
$\mathcal{L}_{\text{source}}^{(2)}(\bm{\rho}(t))$ creates initial
correlations, depending on the phase $\phi$, between the two nodes, while
$\mathcal{L}_{\text{source}}^{(1)}(\bm{\rho}(t))$ does not. 
How these initial
correlations effect the transport properties will be adressed in the
following section.

\section{Transport with and without initial correlations}\label{sec:three}
\begin{figure}
 \begin{center}
  \includegraphics[width=\columnwidth]{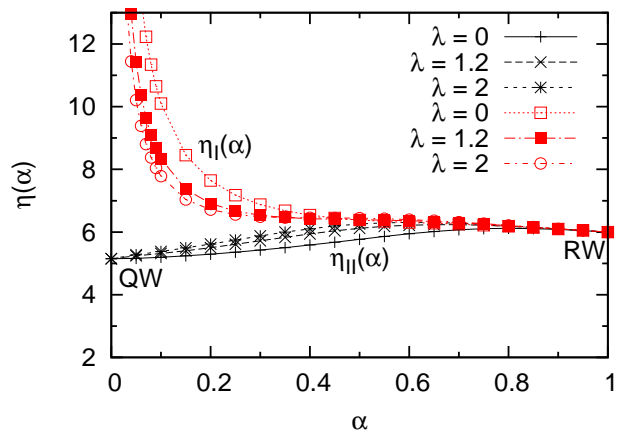}
 \end{center}
 \caption{The dependence on the dephasing rate $\lambda$, with $\Gamma = 0.5$ and $\gamma = 1$. The red curves correspond to 
 an uncorrelated source and the black curves correspond to a correlated source with phase $\phi = 0$.}\label{fig:lambda-dep}
\end{figure}
In previous work we used the expected survival time (EST) $\eta$ as a
measure for the transport properties of the excitation in the network \cite{Schijven2011}.
Here we use it specifically to study the effects of initial correlations
on the transport. This EST is defined as the average time
it needs for the excitation to move completely from the source to the
drain:
\be
  \eta(\alpha) = \int\limits_0^\infty dt \ \left( 1 - \rho_{N+1,N+1}(t, \alpha)
\right).
\ee
The following representation of the EST in terms of the Laplace transforms $\hat{\rho}_{kk}(s)$
of the components of the density matrix, allows for a more convenient way
to obtain analytical expressions for the EST \cite{Schijven2011}:
\be\label{eq:est-laplace}
  \eta(\alpha) = \lim_{s\to 0} \sum_{k=0}^N \hat{\rho}_{kk}(s,\alpha) .
\ee
We denote the EST corresponding to a source feeding independently the two
nodes $1$ and $2$, Eq.
\eqref{eq:uncorrelated}, as $\eta_I(\alpha)$ and the EST corresponding to
the source feeding into an entangled state, Eq. \eqref{eq:correlated}, as
$\eta_{II}(\alpha)$. To illustrate the key effects, we assume, for
simplicity, that $E_1 = E_2 = E_3 = 1$, $V_{13} = V_{23} = 1$, $\gamma =
1$ and $V_{12} = 0$. That is, we focus on the situation when there is no
bond between nodes $1$ and $2$. For these parameters, it follows from Eq. \eqref{eq:est-laplace} that the
ESTs $\eta_I(\alpha)$ and $\eta_{II}(\alpha)$ take the form
\begin{eqnarray}
  \eta_I(\alpha) &=& 1/\Gamma + f(\alpha)/g(\alpha) \label{eq:soleta1}\\
  \eta_{II}(\alpha) &=& 1/\Gamma + [ f(\alpha) - h(\alpha)\cos\phi ]/g(\alpha) \label{eq:soleta2},
\end{eqnarray}
with $h(\alpha) = 4(1-\alpha)^2$ and
\begin{eqnarray}
 f(\alpha) &=& 4 + \alpha(17+13\lambda)+2\alpha^2(\lambda(\lambda-8)-19) \\
           && + 3\alpha^3(11+\lambda(9+2\lambda)) \nonumber \\
 g(\alpha) &=& 4\alpha(2+\lambda) - \alpha^2 (15+7\lambda) + \alpha^3 (11+\lambda(9+2\lambda)). \nonumber
\end{eqnarray}
The dependence of $\eta_{II}(\alpha)$ on the phase $\phi$ is therefore
proportional to $\cos\phi$. Its amplitude $-h(\alpha)/g(\alpha)$ is a
monotonically decreasing function of $\alpha$ and vanishes when $\alpha =
1$. Therefore $\eta_{II}(\alpha)$ converges to $\eta_{I}(\alpha)$ when
$\alpha\to 1$, where they both reach the value $4 + 1/\Gamma$. This also
shows that for $\phi\in[0,\pi/2)$ and $\phi\in (3\pi/2, 2\pi]$,
$\eta_{II}(\alpha) < \eta_I(\alpha)$ and that the converse result holds
for $\phi\in (\pi/2, 3\pi/2)$.  In the limit $\alpha \to 0$ we find that
$\eta_{I}(0) = \infty$, while
\be 
  \lim_{\alpha\to 0}\eta_{II}(\alpha) =
\left\{\begin{array}{cc}\displaystyle 
\frac{1}{\Gamma} + \frac{25+13\lambda}{8+4\lambda}  & \mbox{for} \quad \phi = 0 \\
      \infty & \mbox{for} \quad \phi\neq 0. \end{array}\right. 
\ee

To illustrate these analytical results we show in Fig. \ref{fig:phi-dep},
for $\lambda = 1$, the dependence of the EST $\eta_{II}(\alpha)$ on the
phase $\phi$ and compare it to $\eta_I(\alpha)$. One clearly observes the
$\cos\phi$-dependence of $\eta_{II}(\alpha)$, see left panel, and an
infinite EST for $\alpha = 0$ and $\phi \neq 0$. 
The infinite EST can be
understood by noting that for these values of $\alpha$ and $\phi$
the dark state is  not influenced by the 
drain and is a stationary state of the system, causing both EST's
to diverge when $\phi\neq 0$. For $\phi = 0$ the state $\ket{\psi}$ is
orthogonal to $\ket{D}$, causing the absence of the dark state in
the full dynamics and leading to complete transfer to the drain.  

In Fig. \ref{fig:lambda-dep} we show for $\phi = 0$ the dependence on the
dephasing rate $\lambda$. We observe that for increasing values of
$\lambda$ the EST $\eta_{I}(\alpha)$ decreases, leading to faster
transport. This resembles noise-assisted transport
found in many other systems \cite{Rebentrost2009,Caruso2009}. The
EST $\eta_{II}(\alpha)$, in contrast, increases for larger values of $\lambda$, leading
to slower transport to the drain. 
Thus, here the optimal transport efficiency is obtained in the purely coherent
case ($\alpha = 0$ and $\lambda = 0$). However, $\eta_{II}(\alpha)$ is always
smaller than
$\eta_{I}(\alpha)$.  We further observe that $\eta_{II}(\alpha)$ increases
until a certain $\alpha_c$, after which it follows the curve of
$\eta_I(\alpha)$. The value of $\alpha_c$ becomes smaller with increasing
dephasing rates. This happens because the dephasing process destroys the
coherences between nodes $1$ and $2$. Therefore after this point, the
initial correlations do not significantly influence the transport
properties anymore and $\eta_{II}(\alpha) \approx \eta_I(\alpha)$. When
$\phi=0$, a tractable analytical expression for $\alpha_c$ is possible:
\be
\alpha_c =
\frac{4}{3+2\lambda}\left[\sqrt{\frac{(2+\lambda^3)}{11+9\lambda+2\lambda^2}}-\frac{1}{4}\right].
\ee
Our results show that
initial correlations, induced by
the source feeding the superposition state $\ket{\psi} = (\ket{1} + \ket{2})/\sqrt{2}$, leads
to faster transport than for feeding any other state of the form
$\ket{\psi} = (\ket{1} + e^{i\phi} \ket{2})/\sqrt{2}$ with $\phi \neq 0$, even
in the presence of dephasing. 
Additionally, if $\cos\phi>0$ one always has $\eta_{II}(\alpha) < \eta_I(\alpha)$, 
see Eqs. \eqref{eq:soleta1} and \eqref{eq:soleta2}. Therefore, when initial correlations are present, a smaller 
coupling to the environment (smaller values of $\alpha$) is sufficient to avoid the 
dark state, compared to the situation without initial correlations.

\section{Summary}
In conclusion, we have shown that by connecting a source to the two end nodes of the
V-shaped network it is possible to induce initial correlations between the
coupled nodes and that these initial correlations can overcome the
detrimental effects of the dark state, leading to complete transfer to the
drain. The source can also feed the two end nodes independently, i.e.,
without initial correlations between the two nodes. Then, the dark state
inhibits complete transfer.
When increasing the coupling to the environment, the differences in the
expected survival times between the two types of sourcing processes
diminish. 
We expect that the results obtained here also
hold for larger networks that exhibit invariant subspaces, as for example
described in \cite{Caruso2009} for fully connected networks, in
\cite{Agliari2010} for larger ring-like structures or in \cite{Agliari2011}
for Erd\"os-R\'enyi graphs. Furthermore, 
our results are also related to the study of electrical
currents through a network of two-level systems, since the current is
related to the long-time limit of the time derivative of the EST. 

\acknowledgments{
 We gratefully acknowledge support from the Deutsche Forschungsgemeinschaft (DFG grant MU2925/1-1).
 Furthermore, we thank A. Blumen, A. Anishchenko and L. Lenz for useful discussions.  
}

%  Bibliography file from jabref
% \bibliography{/home/piet/Documents/journal-papers/jabref-database.bib}

\begin{thebibliography}{10}%
\makeatletter
\providecommand \@ifxundefined [1]{%
 \ifx #1\undefined \expandafter \@firstoftwo
 \else \expandafter \@secondoftwo
\fi
}%
\providecommand \@ifnum [1]{%
 \ifnum #1\expandafter \@firstoftwo
 \else \expandafter \@secondoftwo
\fi
}%
\providecommand \enquote [1]{``#1''}%
\providecommand \bibnamefont  [1]{#1}%
\providecommand \bibfnamefont [1]{#1}%
\providecommand \citenamefont [1]{#1}%
\providecommand\href[0]{\@sanitize\@href}%
\providecommand\@href[1]{\endgroup\@@startlink{#1}\endgroup\@@href}%
\providecommand\@@href[1]{#1\@@endlink}%
\providecommand \@sanitize [0]{\begingroup\catcode`\&12\catcode`\#12\relax}%
\@ifxundefined \pdfoutput {\@firstoftwo}{%
 \@ifnum{\z@=\pdfoutput}{\@firstoftwo}{\@secondoftwo}%
}{%
 \providecommand\@@startlink[1]{\leavevmode}%
 \providecommand\@@endlink[0]{}%
}{%
 \providecommand\@@startlink[1]{%
  \leavevmode
  \pdfstartlink
   attr{/Border[0 0 1 ]/H/I/C[0 1 1]}%
   user{/Subtype/Link/A<</Type/Action/S/URI/URI(#1)>>}%
  \relax
 }%
 \providecommand\@@endlink[0]{\pdfendlink}%
}%
\providecommand \url  [0]{\begingroup\@sanitize \@url }%
\providecommand \@url [1]{\endgroup\@href {#1}{\urlprefix}}%
\providecommand \urlprefix [0]{URL }%
\providecommand \Eprint[0]{\href }%
\@ifxundefined \urlstyle {%
  \providecommand \doi [1]{doi:\discretionary{}{}{}#1}%
}{%
  \providecommand \doi [0]{doi:\discretionary{}{}{}\begingroup
  \urlstyle{rm}\Url }%
}%
\providecommand \doibase [0]{http://dx.doi.org/}%
\providecommand \Doi[1]{\href{\doibase#1}}%
\providecommand \bibAnnote [3]{%
  \BibitemShut{#1}%
  \begin{quotation}\noindent
    \textsc{Key:}\ #2\\\textsc{Annotation:}\ #3%
  \end{quotation}%
}%
\providecommand \bibAnnoteFile [2]{%
  \IfFileExists{#2}{\bibAnnote {#1} {#2} {\input{#2}}}{}%
}%
\providecommand \typeout [0]{\immediate \write \m@ne }%
\providecommand \selectlanguage [0]{\@gobble}%
\providecommand \bibinfo [0]{\@secondoftwo}%
\providecommand \bibfield [0]{\@secondoftwo}%
\providecommand \translation [1]{[#1]}%
\providecommand \BibitemOpen[0]{}%
\providecommand \bibitemStop [0]{}%
\providecommand \bibitemNoStop [0]{.\EOS\space}%
\providecommand \EOS [0]{\spacefactor3000\relax}%
\providecommand \BibitemShut [1]{\csname bibitem#1\endcsname}%
%</preamble>
\bibitem{VanKampen1990}%
  \BibitemOpen
  \bibfield{author}{%
  \bibinfo {author} {\bibfnamefont{N.~V.}\ \bibnamefont{Kampen}},\ }%
  \emph{\bibinfo {title} {Stochastic Processes in Physics and Chemistry}}\
  (\bibinfo {publisher} {North Holland, Amsterdam},\ \bibinfo {year} {1990})%
  \bibAnnoteFile{NoStop}{VanKampen1990}%
\bibitem{Farhi1998}%
  \BibitemOpen
  \bibfield{author}{%
  \bibinfo {author} {\bibfnamefont{E.}~\bibnamefont{Farhi}}\ and\ \bibinfo
  {author} {\bibfnamefont{S.}~\bibnamefont{Gutmann}},\ }%
  \bibfield{journal}{%
  \Doi{10.1103/PhysRevA.58.915}{\bibinfo {journal} {Phys. Rev. A}}\ }%
  \textbf{\bibinfo {volume} {58}},\ \bibinfo {pages} {915} (\bibinfo {year} {1998})%
  \bibAnnoteFile{NoStop}{Farhi1998}%
\bibitem{Muelken2011}%
  \BibitemOpen
  \bibfield{author}{%
  \bibinfo {author} {\bibfnamefont{O.}~\bibnamefont{M\"ulken}}\ and\ \bibinfo
  {author} {\bibfnamefont{A.}~\bibnamefont{Blumen}},\ }%
  \bibfield{journal}{%
  \bibinfo {journal} {Phys. Rep.}\ }%
  \textbf{\bibinfo {volume} {502}},\ \bibinfo {pages} {37} (
  \bibinfo {year} {2011})%
  \bibAnnoteFile{NoStop}{Muelken2011}%
\bibitem{BreuerOpenQS}%
  \BibitemOpen
  \bibfield{author}{%
  \bibinfo {author} {\bibfnamefont{H.}~\bibnamefont{Breuer}}\ and\ \bibinfo
  {author} {\bibfnamefont{F.}~\bibnamefont{Petruccione}},\ }%
  \emph{\bibinfo {title} {The theory of open quantum systems}}\ (\bibinfo
  {publisher} {Oxford University Press},\ \bibinfo {year} {2010})%
  \bibAnnoteFile{NoStop}{BreuerOpenQS}%
\bibitem{Whitfield2010}%
  \BibitemOpen
  \bibfield{author}{%
  \bibinfo {author} {\bibfnamefont{J.}~\bibnamefont{Whitfield}}, \bibinfo
  {author} {\bibfnamefont{C.~A.}\ \bibnamefont{Rodriguez-Rosario}},\ and\
  \bibinfo {author} {\bibfnamefont{A.}~\bibnamefont{Aspuru-Guzik}},\ }%
  \bibfield{journal}{%
  \Doi{10.1103/PhysRevA.81.022323}{\bibinfo {journal} {Phys. Rev. A}}\ }%
  \textbf{\bibinfo {volume} {81}},\ \bibinfo {pages} {022323} (
  \bibinfo {year} {2010})%
  \bibAnnoteFile{NoStop}{Whitfield2010}%
\bibitem{Dominguez2011}%
  \BibitemOpen
  \bibfield{author}{%
  \bibinfo {author} {\bibfnamefont{F.}~\bibnamefont{Dom\'inguez}}, \bibinfo
  {author} {\bibfnamefont{S.}~\bibnamefont{Kohler}},\ and\ \bibinfo {author}
  {\bibfnamefont{G.}~\bibnamefont{Platero}},\ }%
  \bibfield{journal}{%
  \Doi{10.1103/PhysRevB.83.235319}{\bibinfo {journal} {\prb}}\ }%
  \textbf{\bibinfo {volume} {83}},\ \bibinfo {eid} {235319} (
  \bibinfo {year} {2011})%
  \bibAnnoteFile{NoStop}{Dominguez2011}%
\bibitem{Michaelis2006}%
  \BibitemOpen
  \bibfield{author}{%
  \bibinfo {author} {\bibfnamefont{B.}~\bibnamefont{Michaelis}}, \bibinfo
  {author} {\bibfnamefont{C.}~\bibnamefont{Emary}},\ and\ \bibinfo {author}
  {\bibfnamefont{C.}~\bibnamefont{Beenakker}},\ }%
  \bibfield{journal}{%
  \Doi{10.1209/epl/i2005-10458-6}{\bibinfo {journal} {Europhys. Lett.}}\ }%
  \textbf{\bibinfo {volume} {73}},\ \bibinfo {pages} {677} (
  \bibinfo {year} {2006})%
  \bibAnnoteFile{NoStop}{Michaelis2006}%
\bibitem{Groth2006}%
  \BibitemOpen
  \bibfield{author}{%
  \bibinfo {author} {\bibfnamefont{C.}~\bibnamefont{Groth}}, \bibinfo {author}
  {\bibfnamefont{B.}~\bibnamefont{Michaelis}},\ and\ \bibinfo {author}
  {\bibfnamefont{C.}~\bibnamefont{Beenakker}},\ }%
  \bibfield{journal}{%
  \Doi{10.1103/PhysRevB.74.125315}{\bibinfo {journal} {Phys. Rev. B.}}\ }%
  \textbf{\bibinfo {volume} {74}},\ \bibinfo {pages} {125315} (
  \bibinfo {year} {2006})%
  \bibAnnoteFile{NoStop}{Groth2006}%
\bibitem{Caruso2009}%
  \BibitemOpen
  \bibfield{author}{%
  \bibinfo {author} {\bibfnamefont{F.}~\bibnamefont{Caruso}}, \bibinfo {author}
  {\bibfnamefont{A.}~\bibnamefont{Chin}}, \bibinfo {author}
  {\bibfnamefont{A.}~\bibnamefont{Datta}}, \bibinfo {author}
  {\bibfnamefont{S.}~\bibnamefont{Huelga}},\ and\ \bibinfo {author}
  {\bibfnamefont{M.}~\bibnamefont{Plenio}},\ }%
  \bibfield{journal}{%
  \Doi{10.1063/1.3223548}{\bibinfo {journal} {J. Chem. Phys.}}\ }%
  \textbf{\bibinfo {volume} {131}},\ \bibinfo {pages} {105106} (
  \bibinfo {year} {2009})%
  \bibAnnoteFile{NoStop}{Caruso2009}%
\bibitem{Sarovar2011}%
  \BibitemOpen
  \bibfield{author}{%
  \bibinfo {author} {\bibfnamefont{M.}~\bibnamefont{Sarovar}}, \bibinfo
  {author} {\bibfnamefont{Y.-C.}\ \bibnamefont{Cheng}},\ and\ \bibinfo {author}
  {\bibfnamefont{K.}~\bibnamefont{Whaley}},\ }%
  \bibfield{journal}{%
  \Doi{10.1103/PhysRevE.83.011906}{\bibinfo {journal} {Phys. Rev. E}}\ }%
  \textbf{\bibinfo {volume} {83}},\ \bibinfo {pages} {011906} (
  \bibinfo {year} {2011})%
  \bibAnnoteFile{NoStop}{Sarovar2011}%
\bibitem{Emary2007}%
  \BibitemOpen
  \bibfield{author}{%
  \bibinfo {author} {\bibfnamefont{C.}~\bibnamefont{Emary}},\ }%
  \bibfield{journal}{%
  \Doi{10.1103/PhysRevB.76.245319}{\bibinfo {journal} {\prb}}\ }%
  \textbf{\bibinfo {volume} {76}},\ \bibinfo {eid} {245319} (
  \bibinfo {year} {2007})%
  \bibAnnoteFile{NoStop}{Emary2007}%
\bibitem{Brandes2005}%
  \BibitemOpen
  \bibfield{author}{%
  \bibinfo {author} {\bibfnamefont{T.}~\bibnamefont{Brandes}},\ }%
  \bibfield{journal}{%
  \Doi{10.1016/j.physrep.2004.12.002}{\bibinfo {journal} {Phys. Rep.}}\ }%
  \textbf{\bibinfo {volume} {408}},\ \bibinfo {pages} {315} (
  \bibinfo {year} {2005})%
  \bibAnnoteFile{NoStop}{Brandes2005}%
\bibitem{Molmer1993}%
  \BibitemOpen
  \bibfield{author}{%
  \bibinfo {author} {\bibfnamefont{K.}~\bibnamefont{M{\o}lmer}}, \bibinfo
  {author} {\bibfnamefont{Y.}~\bibnamefont{Castin}},\ and\ \bibinfo {author}
  {\bibfnamefont{J.}~\bibnamefont{Dalibard}},\ }%
  \bibfield{journal}{%
  \Doi{10.1364/JOSAB.10.000524}{\bibinfo {journal} {J. Opt. Soc. Am. B}}\ }%
  \textbf{\bibinfo {volume} {10}},\ \bibinfo {pages} {524} (
  \bibinfo {year} {1993})%
  \bibAnnoteFile{NoStop}{Molmer1993}%
\bibitem{Agliari2010}%
  \BibitemOpen
  \bibfield{author}{%
  \bibinfo {author} {\bibfnamefont{E.}~\bibnamefont{Agliari}}, \bibinfo
  {author} {\bibfnamefont{O.}~\bibnamefont{M\"ulken}},\ and\ \bibinfo {author}
  {\bibfnamefont{A.}~\bibnamefont{Blumen}},\ }%
  \bibfield{journal}{%
  \bibinfo {journal} {Int. J. Bifurcat. Chaos}\ }%
  \textbf{\bibinfo {volume} {20}},\ \bibinfo {pages} {271} (\bibinfo {year}
  {2010})%
  \bibAnnoteFile{NoStop}{Agliari2010}%
\bibitem{Rebentrost2009}%
  \BibitemOpen
  \bibfield{author}{%
  \bibinfo {author} {\bibfnamefont{P.}~\bibnamefont{Rebentrost}}, \bibinfo
  {author} {\bibfnamefont{M.}~\bibnamefont{Mohseni}}, \bibinfo {author}
  {\bibfnamefont{I.}~\bibnamefont{Kassal}}, \bibinfo {author}
  {\bibfnamefont{S.}~\bibnamefont{Lloyd}},\ and\ \bibinfo {author}
  {\bibfnamefont{A.}~\bibnamefont{Aspuru-Guzik}},\ }%
  \bibfield{journal}{%
  \Doi{10.1088/1367-2630/11/3/033003}{\bibinfo {journal} {New. J. Phys.}}\ }%
  \textbf{\bibinfo {volume} {11}},\ \bibinfo {pages} {033003} (
  \bibinfo {year} {2009})%
  \bibAnnoteFile{NoStop}{Rebentrost2009}%
\bibitem{Schijven2011}%
  \BibitemOpen
  \bibfield{author}{%
  \bibinfo {author} {\bibfnamefont{P.}~\bibnamefont{Schijven}}, \bibinfo
  {author} {\bibfnamefont{J.}~\bibnamefont{Kohlberger}}, \bibinfo {author}
  {\bibfnamefont{A.}~\bibnamefont{Blumen}},\ and\ \bibinfo {author}
  {\bibfnamefont{O.}~\bibnamefont{M\"ulken}}}%
   (\bibinfo {year} {2011}),\
  \Eprint{http://arxiv.org/abs/arXiv:1108.4254}{arXiv:1108.4254}%
  \bibAnnoteFile{NoStop}{Schijven2011}%
\bibitem{Agliari2011}%
  \BibitemOpen
  \bibfield{author}{%
  \bibinfo {author} {\bibfnamefont{E.}~\bibnamefont{Agliari}},\ }%
  \bibfield{journal}{%
  \Doi{10.1016/j.physa.2011.01.021}{\bibinfo {journal} {Physica A}}\ }%
  \textbf{\bibinfo {volume} {390}},\ \bibinfo {pages} {1853} (
  \bibinfo {year} {2011})%
  \bibAnnoteFile{NoStop}{Agliari2011}%
\end{thebibliography}

% Copy-paste from .bbl file 
%Merlin.mbs v4.21 2009-07-09.
%

\end{document}